\documentclass[intlimits,twoside,a4paper]{article}

\usepackage{amsmath,amssymb}
\usepackage{graphicx}

\usepackage[T2A]{fontenc}
\usepackage[cp1251]{inputenc}
\usepackage[eqsecnum]{cmpj2}

\usepackage{textcomp}
\usepackage{here}

\issue{2014}{17}{4}{43302}
\doinumber{10.5488/CMP.17.43302}

\title[Binding of ionic surfactants to flexible polyions]%
{A model study of cooperative binding of ionic surfactants
to oppositely charged flexible polyions}
\author[T. Nishio \textsl{et al.}]{T. Nishio\refaddr{label1},
        T. Shimizu\refaddr{label2}, Sh. Yoshida\refaddr{label1}, A. Minakata\refaddr{label1}}
\addresses{
\addr{label1} Department of Integrated Human Sciences (Physics), Hamamatsu University School of Medicine, \\
Hamamatsu 431-3192, Japan
\addr{label2} Department of Electronic and Information System Engineering, Faculty of Science and Technology, \\
Hirosaki University, Hirosaki 036-8561, Japan
}

\date{Received May 16, 2014, in final form September 26, 2014}
\authorcopyright{T. Nishio, T. Shimizu, Sh. Yoshida,  A. Minakata, 2014}

\begin{document}

\maketitle

\begin{abstract}
A novel statistical model for the cooperative binding of monomeric ligands to a linear lattice is developed to study the interaction of ionic surfactant molecules with flexible polyion chain in dilute solution. Electrostatic binding of a ligand to a site on the polyion and hydrophobic associations between the neighboring bound ligands are assumed to be stochastic processes.   Ligand association separated by several lattice points within defined width is introduced for the flexible polyion.   Model calculations by the Monte Carlo method are carried out to investigate the binding behavior.   The hypothesis on the ligand association and its width on the chain are of importance in determining critical aggregation concentration and binding isotherm.   The results are reasonable for the interpretations of several surfactant-flexible polyion binding experiments.   The implications of the approach are presented and discussed.%
\pacs 36.20.Kd, 61.25.he, 82.60.Hc, 82.70.Uv, 87.10.Mn, 87.16.A-
\end{abstract}

\section{Introduction}

    Complex formation of surfactant molecules with polymer chain is one of the most important and attractive subjects in colloid and polymer science.   It is also useful in various fields of applied chemistry, such as pharmaceutical chemistry, food science, cosmetic manufacture, and so on.   Extensive studies have been carried out on interaction of ionic surfactants with oppositely charged linear polyion, in particular~\cite{Lb301,Lb302,Lb303}.

    One of the interesting subjects to be revealed in this field is the cooperative nature of surfactant binding due to the hydrophobic interactions among bound surfactant molecules on the polyion chain.   To know about it, it is essential to understand the behavior of binding isotherms as well as the critical aggregation concentration (CAC) which characterizes the onset of surfactant binding to polyion \cite{Lb304,Lb305}.   The cooperative nature is due to the side-by-side hydrophobic interactions (association) of the aliphatic tail of the surfactant molecules bound to the polyion chain.

    Concerning the theoretical approaches, simple cooperative site binding theory based on the Ising model has been widely used to successfully analyze the binding isotherms in most cases of dilute solutions of stiff polyions \cite{Lb306,Lb307}.   In our previous works, the effects of steric hindrance and of intrapolymer cooperative interaction between bound ligands across an unoccupied binding site, named skip-binding, are investigated by expanding the simple Ising model.   The effect of steric hindrance due to head group size should be taken into account in some systems \cite{Lb308,Lb310}.   Data fittings using matrix method to this modified binding model show that this approach is available to the analysis of binding isotherms to stiff polyions \cite{Lb309}.   Furthermore, it is important to consider the effect of polyion flexibility on the surfactant binding to understand several experimental studies \cite{Lb311,Lb312,Lb313}.

    However, there seem to be still at least two shortcomings in these models at the moment.   Firstly, the aliphatic tail of the bound surfactant molecules has some degree of freedom with its binding orientation: this should lead the neighboring interactions between the bound surfactants to be stochastic process essentially.   Secondly, the hydrophobic association between the bound surfactants separated by several sites could substantially take place in the ionic surfactant-flexible polyion system.   These two new points of view should be taken into account in the site binding model treatment.

    In the improved model, the polyion molecule, on which bound monomeric surfactant molecules associate with each other, is also assumed to be a linear lattice chain.   Here, electrostatic ligand binding to a lattice point and the hydrophobic association between bound ligands are assumed to be stochastic processes according to the free energy change.   In addition, the associations between bound ligands are acceptable, when their separation is within the given range on the linear lattice.   This expansion of the binding model is necessary for a sufficiently flexible polyion chain.   The calculation procedure using random numbers is employed to obtain the binding isotherms of the system.

    In this paper, the typical calculations of the parameter dependence are attempted for the interpretation of the binding isotherm of ionic surfactants to the flexible polyion with standard Monte Carlo (MC) algorithm.   The results of this simple one-dimensional model calculation show an apparent increase of the surfactant binding, and is reasonable for the interpretation of the characteristics observed in the experimental studies.   The expansion of the scheme on the association of a pair of the ligands is essential for the binding to the flexible polyion.   The present approach for a new lattice model should be feasible to analyze the experimental results of the system of ionic surfactant-flexible polyion.

\section{Theory}

\subsection{The model and parameters}

    In this model, a ligand is not necessary to interact with the ligands on the adjacent binding sites, differently from the ordinary Ising models including our previous ones.   It is based on the assumption that the electrostatic interaction permits the ligand bindings in spatially different orientations around the polyion, although the model does not explicitly include the electrostatic interaction.   Another significant point of the model is the introduction of the association across the multiple sites.   The characteristic parameter association width $n$, mentioned later on, is introduced to the model to reflect the persistence length of a polyion.   Namely, the more flexible the polyion backbone is, the larger $n$ the polyion has.   These assumptions enable the association between ligands, separated by several (occupied and/or non-occupied) binding sites on the polyion.   This brings a partially two- or three-dimensional character into the one-dimensional lattice, and improves the behavior of ligand associations which is more suitable for analysis.

\begin{figure}[htb]
\centerline{\includegraphics[width=0.55\textwidth]{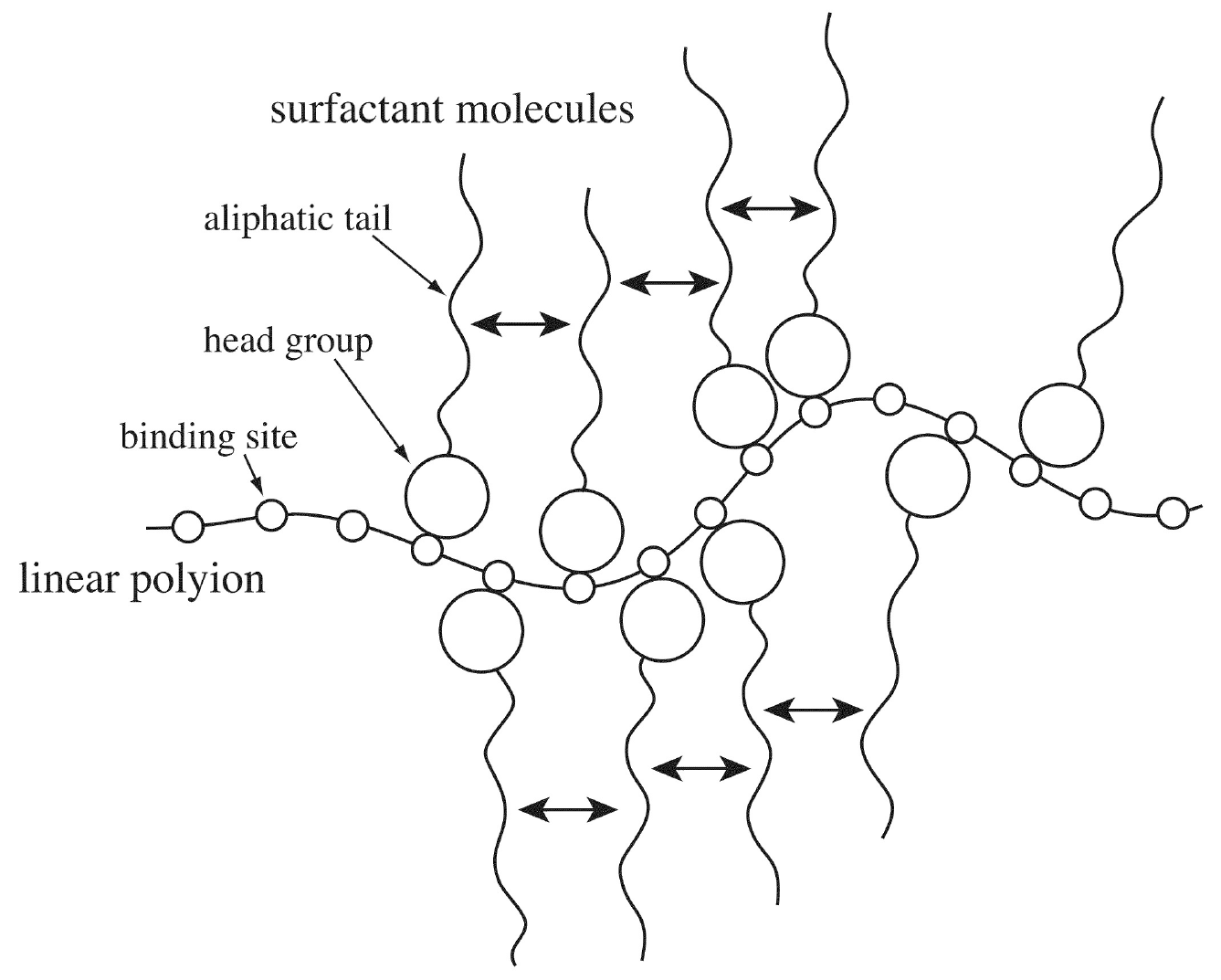}}
\caption{Schematic representation of the surfactant binding to the sites on the flexible polyion in the present model.   The possible pair associations between bound ligands in the case of `association width' $n = 4$ are indicated by thick arrows between aliphatic tails of surfactant molecules.} \label{fig-1}
\end{figure}

    A linear chain consisting of $n$ equivalent charged lattice points (binding sites) is regarded as a model of a flexible polyion.    A monomeric ligand (oppositely charged ionic surfactant molecule) can bind to one of the sites with the binding free energy change $\Delta\epsilon_\mathrm{b}$, where the intrinsic binding free energy change is $\Delta\epsilon^0_\mathrm{b}$.   The main part of the change $\Delta\epsilon^0_\mathrm{b}$  is due to electrostatic energy change between a charged site and an ionic head group of the surfactant.   Bound ligand is capable of stochastically associating with other bound ligands with the association free energy change $\Delta\epsilon_\mathrm{a}$  primarily due to the hydrophobic interactions between the aliphatic tails.   This association is assumed to cover the bound ligand on the $n$-th neighboring lattice points on both sides due to polyion chain collapse.   The free energy change contains the contribution from the polyion deformation.   The number $n$, dependent on polymer configuration, polymer flexibility, and surfactant size, is called `association width' in this manuscript.   In the case of $n \geqslant 2$, the association is possible between the bound ligand separated by lattice point(s).   One ligand can simply associate with two ligands on both sides of the polyion chain in the present model, as shown in figure~\ref{fig-1}.

    The pair association between the bound ligands is chosen to be the nearest neighbor within the association width $n$.   In this model, electrostatic ligand binding without hydrophobic association is permitted even within the association width.   This association does not prevent the associations by other ligands between the pair (see figure~\ref{fig-1}).   In our previous work, cooperative ligand association across only one unoccupied site (skip-binding) was introduced for the surfactant association on the relatively stiff chain \cite{Lb309}.   The assumption in the present model is expanded more suitably for the flexible polyion chain, although the formation of very long loops is ignored and the change of the model makes the analytical calculation difficult.

    The other quantities and the parameters used in this manuscript are defined as, ${C}_\mathrm{f}$, free ligand concentration, $\theta$, binding fraction of sites (degree of binding), $K$, intrinsic binding constant, and $u$, cooperativity parameter where  $\sigma=1/{ u}$ was used as a cooperativity parameter in our previous papers.

    In the present model, five (one ligand-free and four bound) states of a binding site are assumed on each lattice point.   Their symbols and definitions are
\begin{align*}
& \mathrm{f}, & & \text{there is no bound ligand, i.e., the site is ligand-free},\\
& \mathrm{b_0}, & &	\text{there is a bound ligand which does not associate with other ligands},\\
& \mathrm{b_ L}, & &		\text{there is a bound ligand which associates with a ligand on left side},\\
& \mathrm{b_ R}, & & \text{there is a bound ligand which associates with a ligand on right side, and}\\
& \mathrm{b_2},& & \text{there is a bound ligand which associates with two ligands on both sides}.
\end{align*}

\begin{figure}[htb]
\centerline{\includegraphics[width=0.55\textwidth]{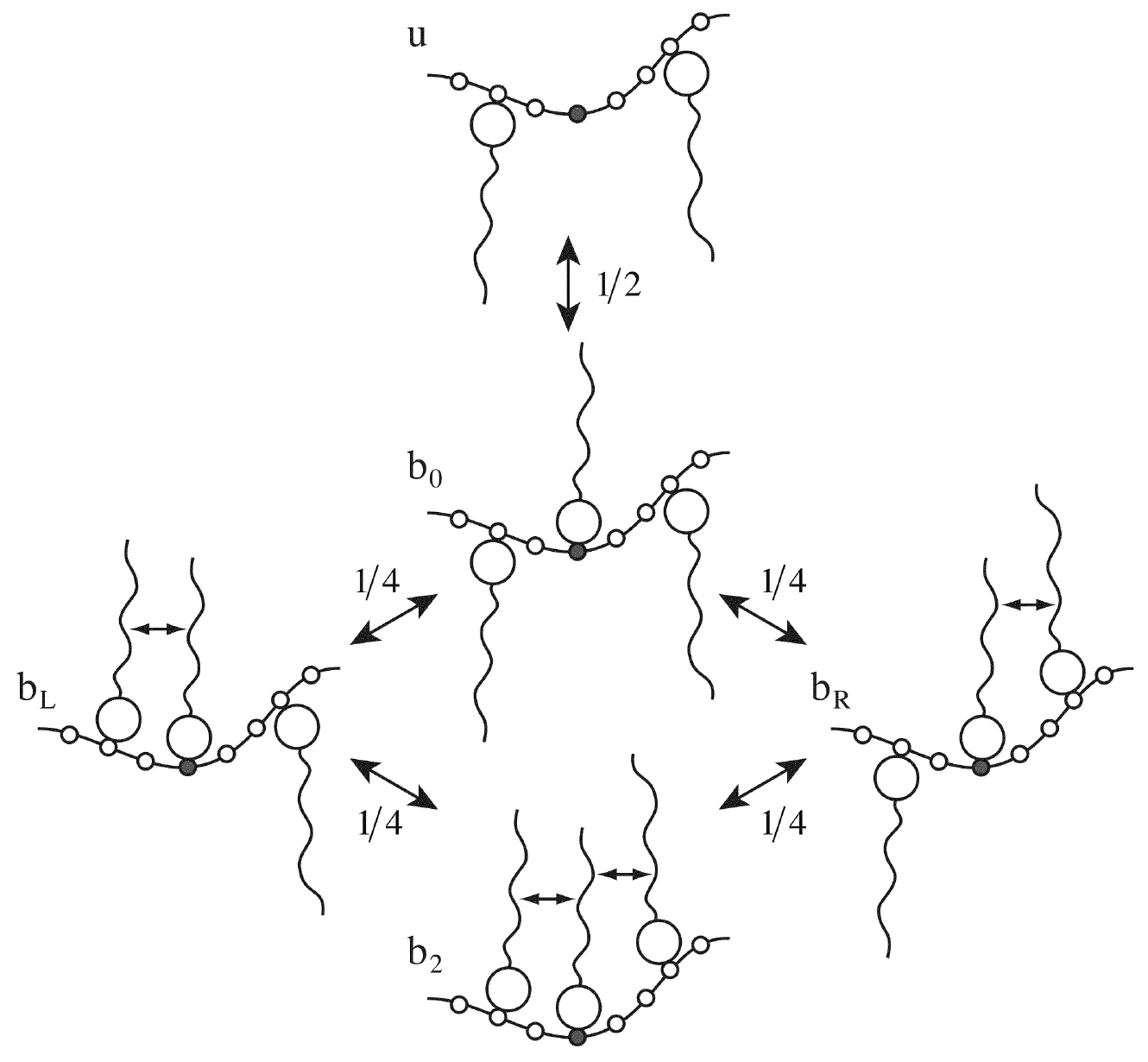}}
\caption{Schematic representation of five states (f, $\mathrm{b_0}$, $\mathrm{b_L}$, $\mathrm{b_R}$ and $\mathrm{b_2}$) of a binding site (small closed circle) on the polyion chain with surfactant and their associations (small arrows).   Large arrows with a fraction indicate the directions of the state transitions and their probabilities of the MC trial, respectively.} \label{fig-2}
\end{figure}

A bound ligand without associated ligands within the width $n$ is in state $\rm b_0$.   If the ligand has associable ligands, its state is stochastically determined according to their energy change in the MC trials.   The states and their transitions are schematically illustrated in figure~\ref{fig-2}.

    The fraction of sum of bound states (binding fraction) $\theta$ is represented as,
\begin{equation}
	\theta = \left[\mathrm{b_0}\right] + \left[\mathrm{b_ L}\right] + \left[\mathrm{b_ R}\right] + \left[\rm b_2\right],
\end{equation}
where the symbol with brackets means their fraction on the polyion and obviously $\left[\mathrm{b_ L}\right] = \left[\mathrm{b_ R}\right]$ since the binding with left-hand and right-hand ligands is equivalent.   The relations of ${K}$ and ${u}$ with these concentrations and the above mentioned free energy changes are expressed as,
\begin{equation}\label{eq-1}
	K = \frac{\left[\rm b_0\right]}{\left[\mathrm{f}\right] C_\mathrm{f}} = \exp\left(-\Delta\epsilon^0_\mathrm{b}/k_\mathrm{B} T\right),
\end{equation}
and
\begin{equation}\label{eq-2}
	u = \frac{1}{\sigma} = \exp\left(-\Delta\epsilon_\mathrm{a}/k_\mathrm{B} T\right),
\end{equation}	
respectively, where ${k}_\mathrm{B}$ is the Boltzmann constant and ${T}$ is absolute temperature.

    The formation of nano- and microparticles is one of the interesting subjects of the surfactant-polyion \cite{Lb314}.   The serial cluster of bound surfactant molecules with the intrapolymer interactions can be treated in the present model.   A series of bound states with associating surfactants on a polyion chain is considered as a `cluster'.   The cluster, whose size (number of associating ligands) is more than one, is serial $\mathrm{b_2}$ states having $\mathrm{b_ L}$ and $\mathrm{b_ R}$ states on both ends, such as ($\mathrm{b_ R\rm b_2\cdots\rm b_2\rm b_ L}$).   The $\mathrm{b_0}$ state without association is also assumed to be the smallest cluster whose size is one.   In figure~\ref{fig-1}, there are three clusters containing an isolated surfactant.   Then, the fraction of the cluster is given by $\left[\rm b_0\right] +\left[\rm b_L\right]$.   The mean size of the cluster $l_\mathrm{C}$ is defined as,
\begin{equation}\label{eq-3}
	l_\mathrm{C} = \frac{\theta}{\left[\mathrm{b_ 0}\right]+\left[\mathrm{b_ L}\right]},
\end{equation}	
where  $l_\mathrm{C} \geqslant 1$.   By a simple idea, at a full binding state   $\theta = 1$, the following relation is assumed
\begin{equation}\label{eq-4}
	u = \frac{\left[\mathrm{b_ L}\right]}{\left[\mathrm{b_ 0}\right]}= \frac{\left[\mathrm{b_ R}\right]}{\left[\mathrm{b_ 0}\right]}= \frac{\left[\mathrm{b_ 2}\right]}{\left[\mathrm{b_ L}\right]}= \frac{\left[\mathrm{b_ 2}\right]}{\left[\mathrm{b_ R}\right]}\,.	\nonumber
\end{equation}
It is expected that the  $l_\mathrm{C}$  value tends to $u + 1$ when $\theta$  approaches unity.   The distribution of the cluster size can be obtained as well.

\subsection{Model calculation}

    In the MC process, the state transition of each site is decided by the probability function with the change of its energy using a pseudo-random number (figure~\ref{fig-2}).   The free energy change of only ligand binding without association between bound ligands ($\mathrm{f} \to \mathrm{b_ 0}$) is
\begin{equation}\label{eq-5}
	\Delta\epsilon_\mathrm{b} = \Delta\epsilon^0_\mathrm{b} - k_\mathrm{B} T \ln\left(C_\mathrm{f}\right) = - k_\mathrm{B} T \ln\left(KC_\mathrm{f}\right),
\end{equation}	
which is dependent on the concentration of free ligand.   The free energy change with association between a pair of bound ligands ($\mathrm{b_ 0} \to \mathrm{b_ L}$ or $\mathrm{b_ L}$ and $\mathrm{b_ L}$ or $\mathrm{b_ R} \to \mathrm{b_ 2}$) is
\begin{equation}\label{eq-6}
	\Delta\epsilon_\mathrm{a} = - k_\mathrm{B} T \ln u.
\end{equation}	
The acceptance of state change trials is determined according to the standard Metropolis algorithm using the probability function of the energy change of the system \cite{Lb315,Lb316}.

    For example, if the state of the site is $\mathrm{b_ 0}$, the probabilities of the MC trial for the transition to the states $\mathrm{f}$, $\mathrm{b_ L}$ and $\mathrm{b_ R}$ are 1/2, 1/4, and 1/4, respectively.   In the process from $\mathrm{b_ 0}$ to $\mathrm{b_ L}$, if the bound ligand which can associate to the objective ligand in the left-hand side within width $n$ is found, the state transition is decided in the above mentioned manner.   If no associable ligand is found, the process is discarded.   Other probabilities of the process and the manner of decision are similarly given in order to maintain the detailed balance (see figure~\ref{fig-2}).

    The number of MC step per site is chosen within the range from $10^5$ to $10^6$ at a given $KC_\mathrm{f}$  ($\Delta\epsilon_\mathrm{b}$ ) value so that the convergence of the quantities is achieved in the early stage (mostly within a few percent of half MC steps).   After half discarded MC steps, the quantities are finally evaluated by summing up in the later MC steps.

    All f state is usually adopted as an initial state of sites for a series of calculations.   A final state at each $KC_\mathrm{f}$ value is used as an initial state of the next point.   The MC trials are carried out in the order of the lattice site number within a single loop.   The results are unaffected even if the site is randomly chosen, or an initial state is changed.

    In the standard calculations, the lattice has $N = 1000$ sites under the periodic boundary condition.   When $N = 5000$, no difference of the result is observed.   The generalized Fibonacci method is used as a random number generator \cite{Lb317}.

\subsection{Analytical solutions}

    In the classical cooperative model without various more realistic effect, we have the well-known equation of binding fraction by Satake and Yang \cite{Lb307}, expressed as,
\begin{equation}\label{eq-7}
	\theta = \frac{1}{2}\left[1-\frac{1-s}{\sqrt{(1-s)^2+4\sigma s}}\right] = \frac{1}{2}\left[1-\frac{1-KuC_\mathrm{f}}{\sqrt{(1-KuC_\mathrm{f})^2+4KC_\mathrm{f}}}\right],
\end{equation}
where well-used parameter $s$ is defined as, $s=KuC_\mathrm{f}$ \cite{Lb307}.   When $u = 1$, the relation means simple binding, as
\begin{equation}\label{eq-8}
	\theta = \frac{KC_\mathrm{f}}{1+KC_\mathrm{f}}\,.
\end{equation}

    The binding isotherm by our previous scheme with a matrix method is also presented in the case of $n = 2$ in figure~\ref{fig-4} and figure~\ref{fig-5} \cite{Lb309}.

\section{Results and discussion}

\subsection{Dependence on the cooperativity parameter $u$}

\begin{figure}[!t]
\centerline{\includegraphics[width=0.55\textwidth]{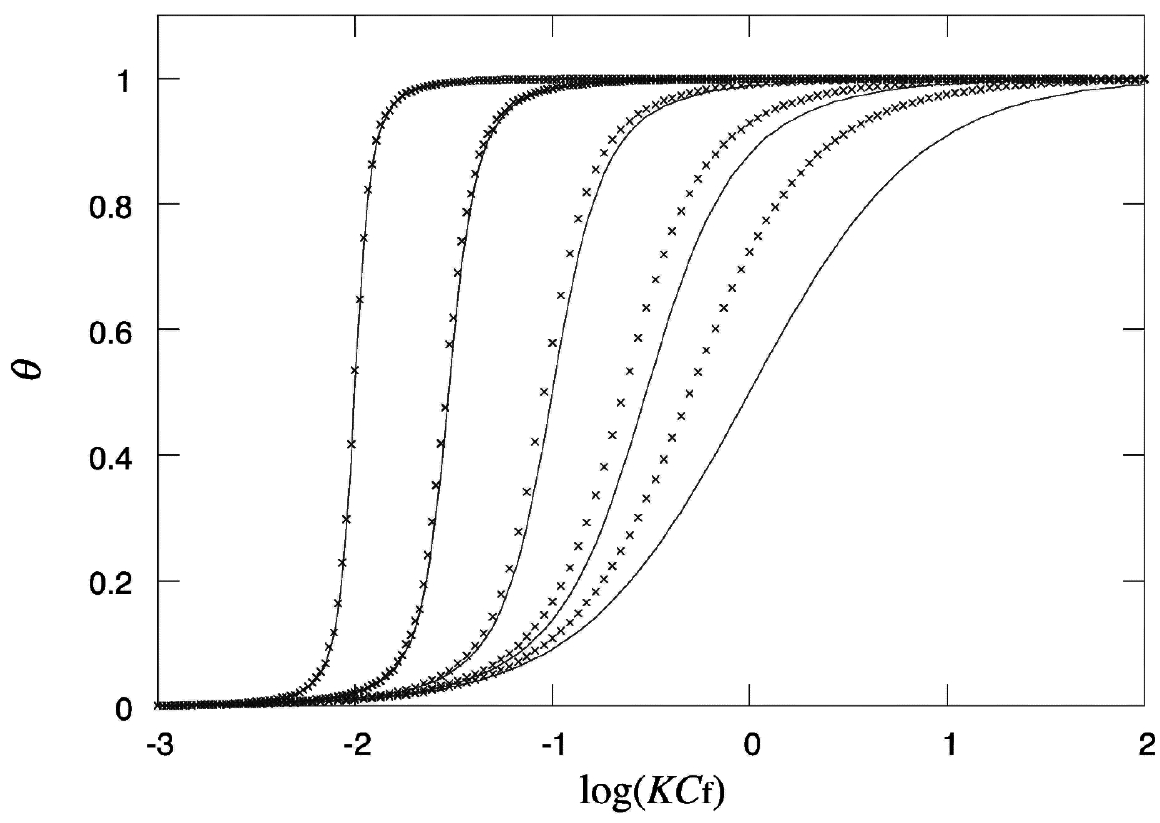}}
\caption{Dependence of binding isotherm on the cooperativity parameter $u$ at association width $n = 1$.   Symbols, MC calculations, lines, curves by the Satake-Yang equation, equation (\ref{eq-7}).   $u$; 1, 3.33 ($1/u = 0.3$), 10, 33.3 ($1/u = 0.03$), and 100, from right to left, respectively.   $N = 1000$ under the periodic boundary condition.} \label{fig-3}
\end{figure}

    First, the fundamental properties of our model system are presented as compared with the classical cooperative bindings of equation~(\ref{eq-7}).  The dependence of binding isotherm on the cooperativity parameter $u$ is presented at the association width $n = 1$ in figure~\ref{fig-3}.   In this case, the association with the nearest neighbor is only considered.   At $u = 1$, the MC calculation curve shows a higher affinity (lower CAC) and cooperativity than the simple binding curve of equation~(\ref{eq-8}) due to the present binding scheme with four bound states.   With an increase of the parameter $u$, the difference between both curves becomes smaller.   At $u = 100$, the differences of  $\theta$ values are within $±10 \%$ at most points.

\begin{figure}[!t]
\centerline{\includegraphics[width=0.55\textwidth]{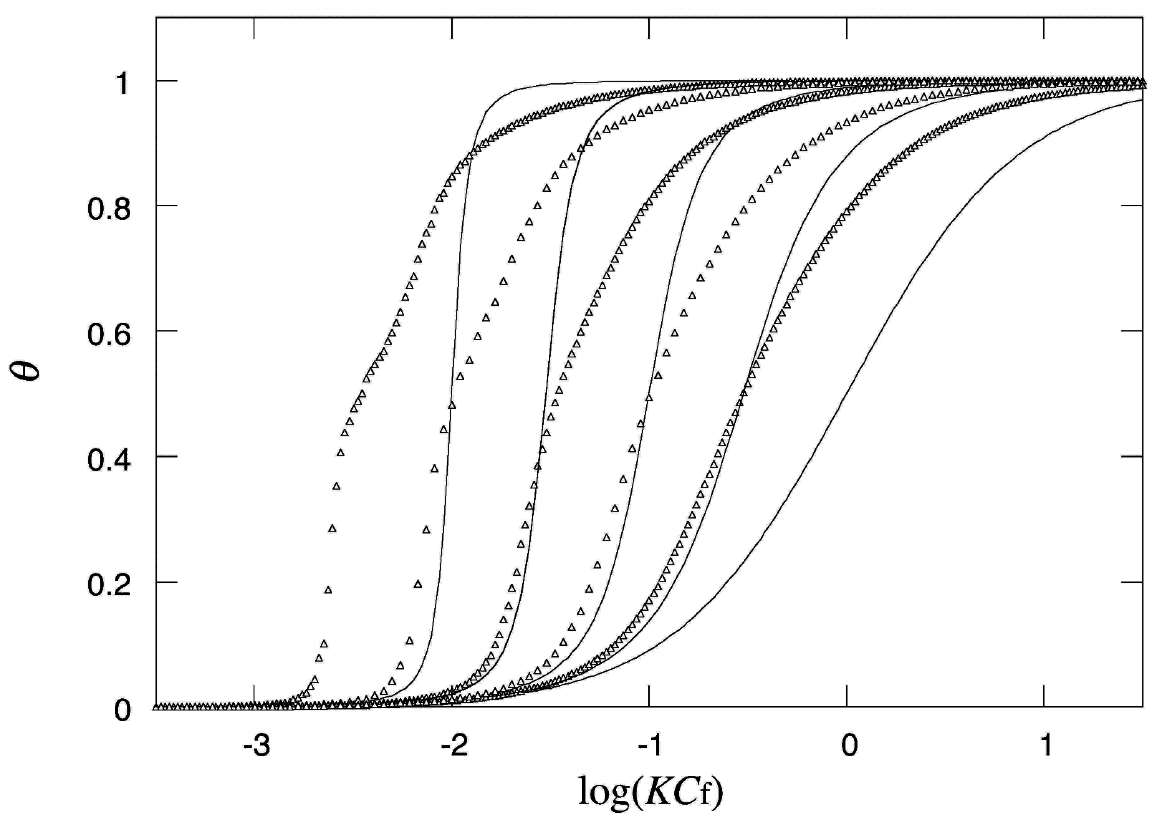}}
\caption{The dependence of binding isotherm on the cooperativity parameter $u$ at association width $n = 4$.   the others are the same as in figure~\ref{fig-3}.} \label{fig-4}
\end{figure}

    In  figure~\ref{fig-4}, the dependence on the cooperativity parameter $u$ is shown in the case of $n = 4$.   Associations between the surfactants separated by binding sites within the width $n = 4$ reduce the CAC values to about one third of the classical model.   The association with larger $u$ value results in higher affinity and asymmetrical binding isotherms in the case of the ligand association across the binding sites.   In the high $u$ cases, most of the bound surfactants provide associations with the neighboring ones in the range $\theta < 0.5$.   Then, the bindings become relatively weak in high $\theta$ range.   At $u = 100$, the MC result shows a slight two-phase binding curve with a shoulder at around  $\theta = 0.5$, corresponding to the rearrangement of clusters (shown below).

    The binding isotherm with a sharp rise and gradual saturation are frequently observed in the experiments of surfactant binding to a flexible polyion\cite{Lb304, Lb318}.   A factor deforming binding isotherm is exhibited from our model calculations.
    In the previous works, we showed that the two-phase binding isotherm is interpreted by the hypothesis of ligand association across a site and/or of steric hindrance.   In the present model, we can observe the similar feature only in the restricted ($n = 2 - 8$ and high $u$) cases.

\subsection{Dependence on the association width $n$}

\begin{figure}[!t]
\centerline{\includegraphics[width=0.55\textwidth]{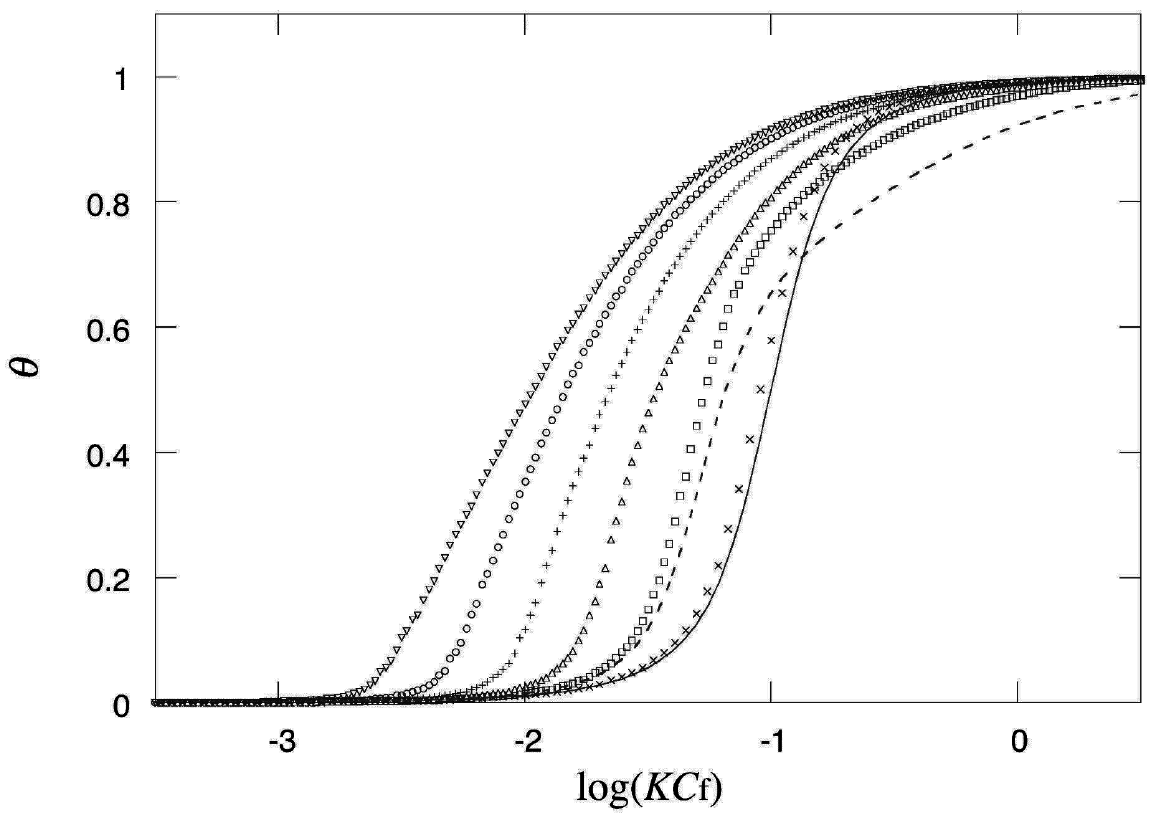}}
\caption{The dependence of binding isotherm on the association width $n$ at cooperativity parameter $u = 10$.   $n$; 1 ($\times$), 2 (square), 4 (triangle), 8 (cross), 16 (circle), and 32 (inverted triangle), from right to left, respectively.   $N = 1000$ under the periodic boundary condition.   Full line is drawn by the Satake-Yang's theory.   Broken line indicates the isotherm of $n = 2$ case calculated in our previous scheme (${\beta}_{1 \rm a}={\beta}_{1 \rm b}={\beta}_1=1$, ${\gamma}=1$, ${\sigma}=1/u=0.1$) \cite{Lb309}.} \label{fig-5}
\end{figure}

    Dependence of the binding isotherm on the association width $n$ is shown in figure~\ref{fig-5} in the case of $u = 10$.   With an increase of the association width $n$, the binding becomes stronger and the curve clearly shows non-symmetrical feature with a sharp rise and relatively gradual saturation.   In the cases of $n = 2$ or 4, cooperative binding looks steep in low $\theta$ range, but in larger $n$ ($n \geqslant  8$) cases, the cooperativity of the isotherm becomes low.

\begin{figure}[!t]
\centerline{\includegraphics[width=0.55\textwidth]{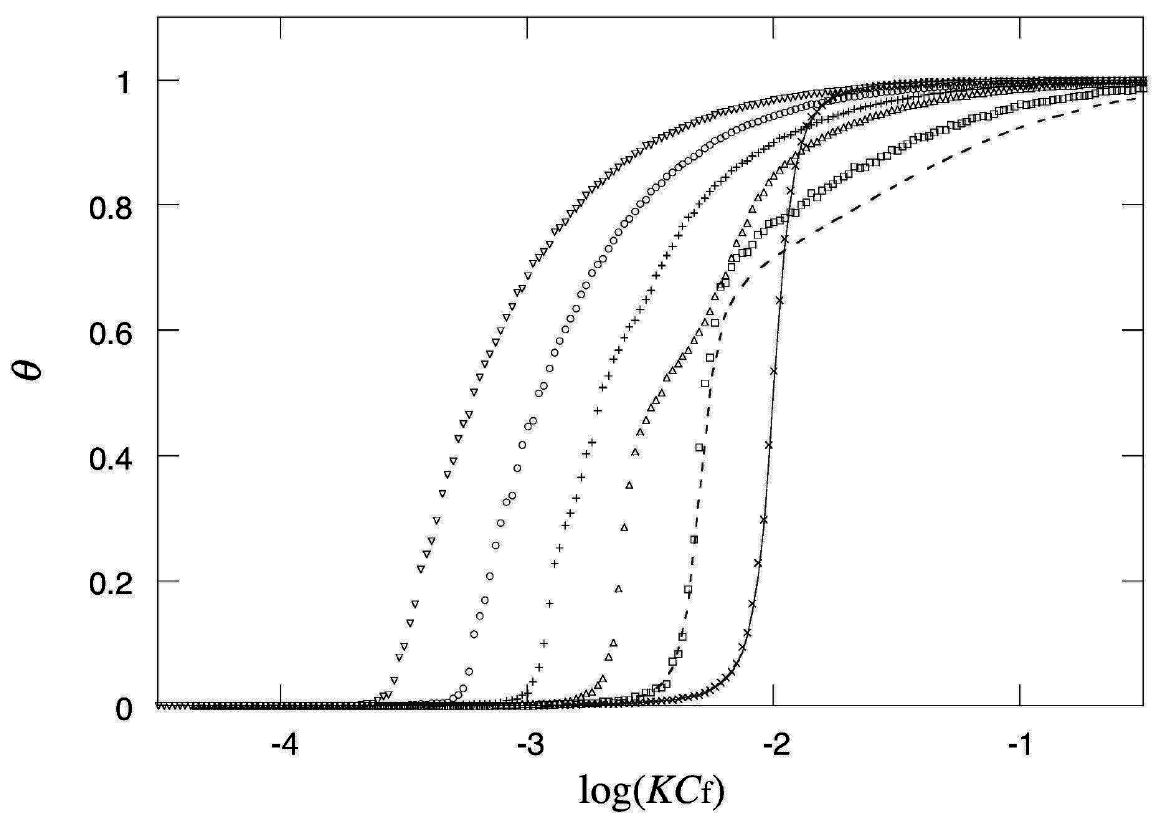}}
\caption{The dependence of binding isotherm on the association width $n$ at $u = 100$.   The others are the same as in figure~\ref{fig-5}.   Broken line indicates the isotherm of $n = 2$ case calculated in our previous scheme (${\beta}_{1 \rm a}={\beta}_{1 \rm b}={\beta}_1=1$, ${\gamma}=1$, ${\sigma}=1/u=0.01$).} \label{fig-6}
\end{figure}

    In a more cooperative case of $u = 100$, the dependence of the curves on $n$ is complicated as shown in figure~\ref{fig-6}.   The distances between the binding isotherms are slightly wider than those in figure~\ref{fig-5}.   When $n = 2$ or 4, a strong two-phase feature appears with a shoulder on the binding isotherm.   In large $n$ ($n \geqslant 8$) cases, the binding isotherm becomes monotonous with a very sharp rise.

    The shoulders of binding isotherms in figure~\ref{fig-4} and figure~\ref{fig-6} are interpreted as follows.   In the cases of $n = 2 - 4$ at high cooperativity parameter for low binding fraction ($\theta$ < 0.5), the ligands tend to entropically bind at intervals, leading to associations across unoccupied site(s).   When the binding fraction becomes around 0.5, the ligands begin to bind to the remaining unoccupied sites.   However, these bindings are suppressed because the newly bound ligand cannot sufficiently associate with the neighboring ligands.   In the higher binding range, recombination of the associations occurs, represented by complicated cluster size changes shown in figure~\ref{fig-8}.   Such a transition of the situation of the ligands makes a shoulder on the binding isotherms.   In the cases of $n = 5 - 8$ at high cooperativity parameter, the binding isotherm frequently shows a stepwise increase of $\theta$.   In the cases of still larger $n$, the transition becomes obscure due to the extensive possibilities of association formations.

    The isotherms in our previous scheme can be calculated only in the $n = 2$ case using the matrix method (broken line in figure~\ref{fig-5} and figure~\ref{fig-6}).   Although the curves and CAC values resemble those of  $n = 2$ in the model, their rising tendency is more gradual in high $\theta$ range than those of the present model due to the incorporation of the flexibility in the present model.

    It has been observed that the long tail of the surfactant molecule enhances the binding affinity, and leads to low CAC \cite{Lb319, Lb305, Lb318}.   In these cases, the long surfactant may induce an increase in association region (association width $n$ in the model).   This can effect the shape of the binding isotherm as well as the CAC value.   The effect of molecular geometry of surfactant must be reconsidered from this point of view~\cite{Lb320,Lb311}.

\subsection{End effect (Dependence on the lattice length)}

\begin{figure}[!h]
\centerline{\includegraphics[width=0.55\textwidth]{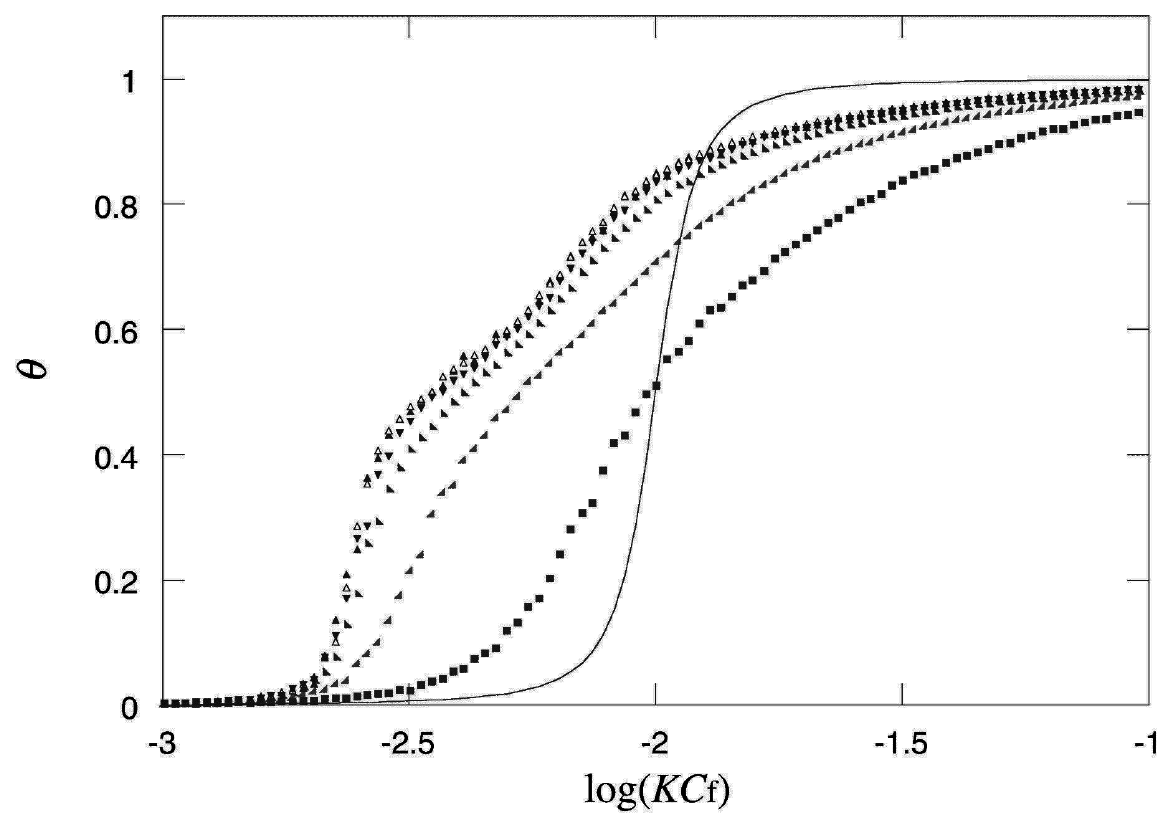}}
\caption{The dependence of binding isotherm on the number of lattice points $N$ at $u = 100$ and $n = 4$ without periodic boundary condition.   $N$; (closed symbols) 1000 (triangle), 300 (inverted triangle), 100 (right-hand triangle, dark), 30 (right-hand triangle, light), and 10 (square), from left to right, respectively.   Open triangle; $N = 1000$ case under periodic boundary condition.} \label{fig-7}
\end{figure}

    The above MC results  are obtained under the periodic boundary condition of $N = 1000$, i.e., by connecting both ends of the lattice.   The end effect and the dependence of the polyion (lattice) length are examined without the periodic boundary condition.  Figure~\ref{fig-7} shows a typical effect of small $N$ values on the isotherms under non-periodic condition at $u = 100$ and $n = 4$, compared with $N = 1000$ case under the periodic boundary condition.

    In the case of $N = 1000$, the end effect cannot be detected, and in the cases of $N \geqslant 100$ the effect is relatively small.   In $N < 100$ cases, however, the shift of the binding curve is significant.   The decrease in
$N$ makes the binding curve smooth and gradual. The end effect is larger in larger $n$ and/or larger $u$ cases (data not shown).

    The effect of polyion size on the surfactant-polyion system has been investigated by several research groups \cite{Lb321,Lb322,Lb323,Lb324,Lb325}.   With a decrease of the polymer chain length, an increase of CAC and a decrease of cooperativity have been observed \cite{Lb321,Lb322}.   The reduction of the binding with a decrease of the polyion length can be interpreted by this calculation.

\subsection{Cluster size}

\begin{figure}[!t]
\centerline{\includegraphics[width=0.55\textwidth]{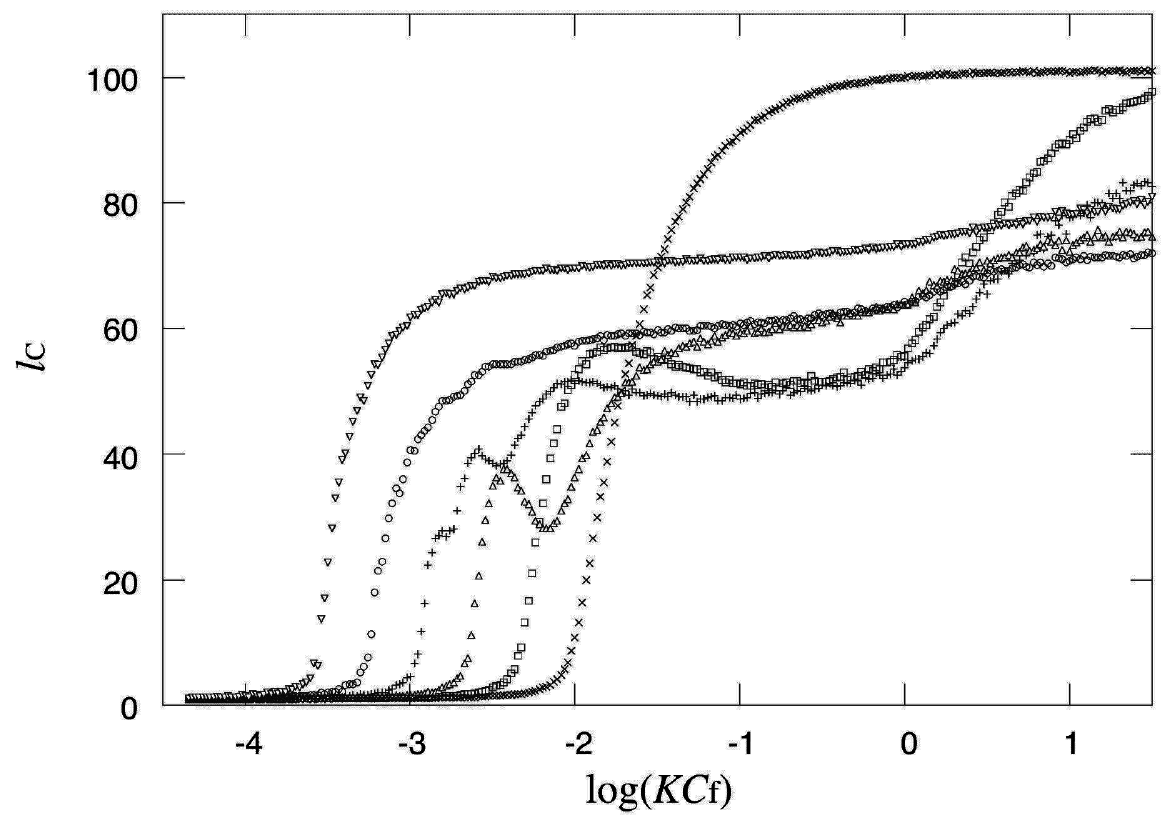}}
\caption{The dependence of variation in mean cluster size $l_\mathrm{c}$ on the association width $n$ at cooperativity parameter $u = 100$.   Symbols are the same as in figure~\ref{fig-6}.} \label{fig-8}
\end{figure}

    In figure~\ref{fig-8}, the changes of mean cluster size $l_\mathrm{c}$  with the surfactant binding are shown with the same parameters as in figure~\ref{fig-6}.   The onset of self-assembly is coincident with the CAC depending on $n$ value.   In cases of $n = 2$, 4, and 8, the curves show up and down in a complex manner.   The cluster size distributions are monotonous due to the one-dimensional lattice system (data not shown).   In cases of $n = 2 - 8$, ligands separated on the lattice induce aggregations in low $\theta$ range with collapse of the polyion chain.   The rearrangements of the association between ligands occur with an increase of the ligand binding in the middle $\theta$ range.   In cases of larger $n$, the rearrangement is more gradual and spread to wider $\theta$ range.   This unexpected behavior may be important to understand the binding of the surfactant to the flexible polyion.

    In the model calculation, large $u$ value leads to high  $l_\mathrm{c}$.   It has been shown that a large cluster size (aggregation number) corresponds to a high cooperativity \cite{Lb326}.   The dependence of the cooperativity on the charge density of the polyion may be interpreted in the present idea.

    In the present model, one ligand can simply associate with two ligands on both sides of the polyion chain.   More complicated models showing phase transition, should be suitable for the coil-globule transition by introducing additional hypotheses \cite{Lb327,Lb328}.

\subsection{Comparison with the experiment}

    Shimizu reported the bindings of cationic surfactants by several flexible poly(carboxylic acid)s using a potentiometric technique \cite{Lb329}.   Most of the binding isotherms show a sharp onset followed by gradual saturation.   The gradual slope in a range of higher $\theta$ means that highly ligand association suppresses a new binding, although the interpretation due to the electrostatic factor is also likely.

\begin{figure}[!b]
\centerline{\includegraphics[width=0.55\textwidth]{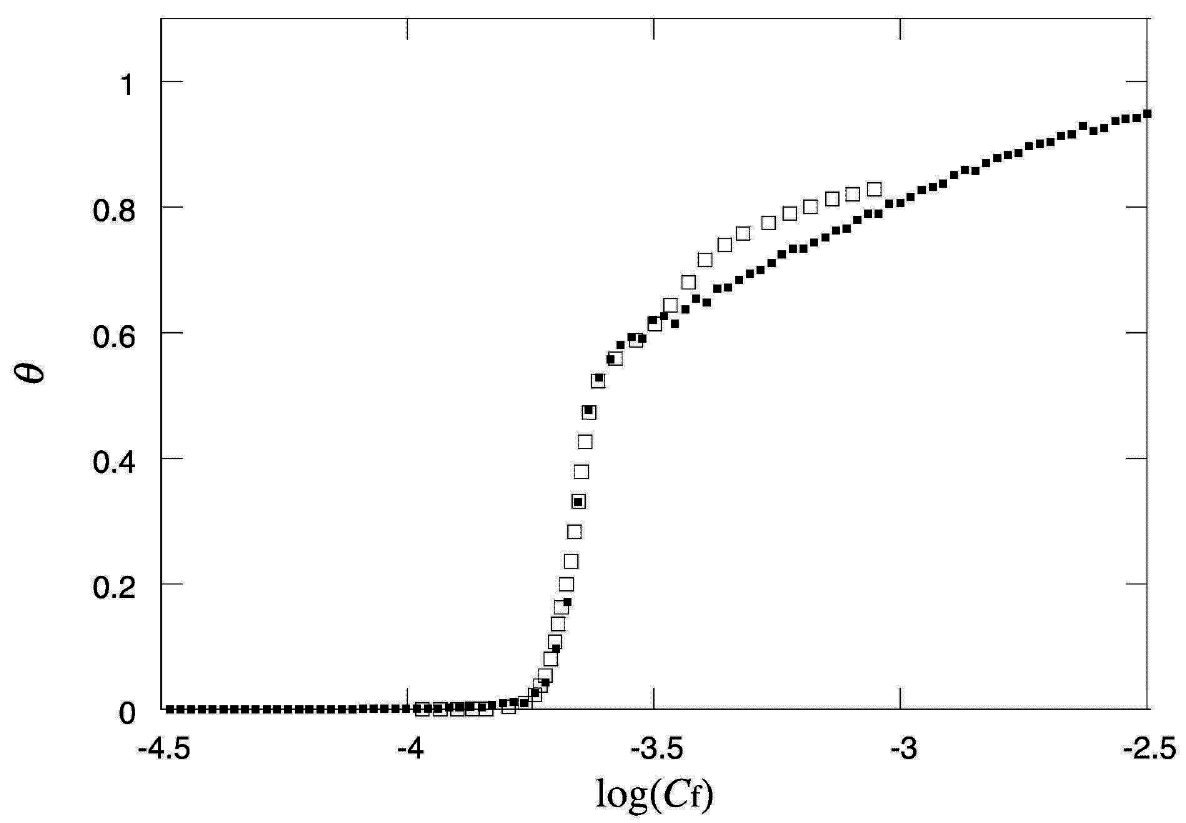}}
\caption{Comparison of the experimental isotherm of dodecylpyridinium ion to poly acrylic acid for $\alpha = 1.0$ (large squares) by Shimizu \cite{Lb329} with the correspondent calculated values (small filled squares).   Experimental conditions; $T = 30$~\textcelsius, $m_\mathrm{NaCl} = 0.01 \ \mathrm{mol} \cdot \mathrm{kg}^{-1}$ H$_2$O, and $m_\mathrm{p} = 0.915$~mN.   Parameters of the model calculation; $N = 1000$, $n = 3$, $\sigma = 1/u = 0.003$, and $\log K = 0.66$ ($\Delta\epsilon^0_\mathrm{b} = -3.8$~kJ/mol, $\Delta\epsilon_\mathrm{a} = -14.6$~kJ/mol).} \label{fig-9}
\end{figure}

    The tentative comparison of the calculated binding curve with an experimental one is executable, although the precise data-fitting in the whole range is difficult.   Figure~\ref{fig-9} shows the binding isotherm of the dodecylpyridinium ion (DP+) by poly acrylic acid (PAA) for $\alpha = 1.0$ at 0.01~m NaCl \cite{Lb329} with the correspondent calculated curve.   In the low surfactant concentration range, reasonable agreement is achieved with the association width $n = 3$ and estimated cooperativity parameter $\sigma = 1/u = 0.003$.
The high cooperativity and the association within the third neighboring ligands lead to this binding profile.
The slight shoulder of the experimental isotherm in the higher range may be due to additional weak associations of the ligands, as observed  in other isotherms \cite{Lb329}.

\section{Conclusions}

    The objective of the study is to analyze the essential characteristics of the system using a simple model.   It seems more suitable rather than molecular simulation studies.   The present model is examined to investigate the binding mechanism of ionic surfactants to a flexible polyion in dilute solution by the MC calculation.   Introducing the non-cooperative binding and the ligand association separated by multiple lattice points, we can show their effects as new factors that determine the binding isotherm.   Charges on the polyion should play an appreciable role in increasing the local concentration of the surfactant molecules.

    In this study, a simple model is examined by the MC calculation to investigate the binding mechanism of ionic surfactants to a flexible polyion in dilute solution.   Charges on the polyion play a role in increasing the local concentration of the surfactant molecules.   Introducing the ligand association separated by lattice points within the association width $n$ in the stochastic processes, we can show its effect as a new factor that determines the binding isotherm.

    In the case at high $u$ of $n = 1$, the results of the present model converge to those of a classical cooperative model.   With making the width $n$ double, the binding affinity looks approximately twofold, i.e., reducing CAC by half.     In the range of $n = 2 - 8$, two-phase or stepwise nature of the isotherm is observed due to the complicated rearrangement of the surfactant cluster on the polyion chain, in particular, in the high $u$ cases.   The short polyion chain also lowers the affinity and their cooperativity in the model due to the end effect.

    The model of cluster formation is so simple that the results are rather limited at the present stage.   However, our approach suggests the essential points of the effect of polyion flexibility on the binding isotherm reflecting the freedom of ligand binding and of their association.

\section*{Acknowledgements}

The calculations in this study were carried out using the workstation in the information technology center of the Hamamatsu University School of Medicine.



\ukrainianpart

\title
{Модельне дослідження колективного зв'язування  іонних сурфактантів з протилежно зарядженими \\ гнучкими полііонами}
\author{Т. Нішіо\refaddr{label1},
        Т. Шімізу\refaddr{label2}, Ш. Йошіда\refaddr{label1}, А. Мінакати\refaddr{label1}}
\addresses{
\addr{label1} Відділ об'єднаних наук про людину (фізика), Медична школа університету  м. Хамацу, \\ Хамацу 431-3192, Японія
\addr{label2} Відділ інженерії електронних та інформаційних систем, факультет природничих наук і технології, \\ Університет м. Хіросакі, Хіросакі 036-8561, Японія
}

\makeukrtitle

\begin{abstract}
\tolerance=3000%
Для вивчення взаємодії молекул іонного сурфактанта з гнучким полііонним ланцюжком у розведеному розчині запропоновано
нову статистичну модель для колективного зв'язування мономерних ліганд з лінійною граткою.
Припускається, що електростатичне зв'язування  ліганди з вузлом на полііоні та гідрофобні  зв'язки між сусіднімі  зв'язаними лігандами
є стохастичними процесами. Для  гнучкого полііона вводиться асоціація ліганд, відокремлених декількома гратковими точками в межах
визначеної ширини. Для того, щоб дослідити поведінку    зв'язуван\-ня, здійсноються обчислення методом Монте Карло.
Гіпотеза асоціації ліганди та її ширини на ланцюжку є важливою для визначення критичної концентрації агрегації та ізотерми зв'язування.
Результати є прийнятними для інтерпретиції декількох експериментів по зв'язуванню  сурфактант-гнучкий полііон.
Представлено та обговорено застосування  методу.%
\keywords взаємодія іонний сурфактант-гнучкий полііон,  гратка лінійного полііона, колективне  зв'язування ліганд,
ізотерми зв'язування, ширина асоціації, модельне обчислення
\end{abstract}

\end{document}